\documentclass{article}
\usepackage[utf8]{inputenc}
\usepackage[utf8]{inputenc}
\usepackage[letterpaper,top=2cm,bottom=2cm,left=3cm,right=3cm,marginparwidth=1.75cm]{geometry}
\usepackage{natbib}
\usepackage{amsmath, amssymb, amsfonts}
\usepackage{graphicx}
\usepackage{xcolor}
\usepackage{bm}
\usepackage[colorlinks=true, allcolors=blue]{hyperref}
\usepackage[affil-it]{authblk}
\usepackage{xcolor}
\newcommand\shorthandon{\catcode`@=\active}
\shorthandon
\def@#1@{\textcolor{cyan}{#1}}%

\title{MOSAIK: Multi-Origin Spatial Transcriptomics Analysis and Integration Kit}

\author[1,2]{Anthony Baptista$^*$}
\author[1]{Rosamond Nuamah}
\author[3,4]{Ciro Chiappini}
\author[1]{Anita Grigoriadis}

\affil[1]{Cancer Bioinformatics, School of Cancer and Pharmaceutical Sciences, Faculty of Life Sciences and Medicine, King’s College London, London, WC2R 2LS, UK}
\affil[2]{The Alan Turing Institute, The British Library, London, NW1 2DB, United Kingdom}
\affil[3]{Centre for Cranio facial and Regenerative Biology, King’s College London, London, SE1 9RT, UK}
\affil[4]{London Centre for Nanotechnology, King’s College London, London WC2R 2LS, UK}

\date{ }

\begin{document}

\maketitle

\section*{Summary}
\hfill

Spatial transcriptomics (ST) has revolutionised transcriptomics analysis by preserving tissue architecture, allowing researchers to study gene expression in its native spatial context. However, despite its potential, ST still faces significant technical challenges. Two major issues include: (1) the integration of raw data into coherent and reproducible analysis workflows, and (2) the accurate assignment of transcripts to individual cells. To address these challenges, we present MOSAIK, the first fully integrated, end-to-end workflow that supports raw data from both NanoString CosMx Spatial Molecular Imager (CosMx) and 10x Genomics Xenium In Situ (Xenium). MOSAIK (Multi-Origin Spatial Transcriptomics Analysis and Integration Kit) unifies transcriptomics and imaging data into a single Python object based on the spatialdata format. This unified structure ensures compatibility with a broad range of Python tools, enabling robust quality control and downstream analyses. With MOSAIK, users can perform advanced analyses such as re-segmentation (to more accurately assign transcripts to individual cells), cell typing, tissue domain identification, and cell-cell communication within a seamless and reproducible Python environment.

\section*{Statement of need}
\hfill

Spatial transcriptomics (ST) enables the study of transcriptomes within intact tissues, which is essential for understanding a cell’s position relative to its neighbours and the surrounding extracellular structures. This spatial context provides crucial insights into cellular phenotype, function, and disease progression, particularly in cancer, where the tumour micro-environment (TME) influences processes such as chemo-resistance \cite{mehraj_tumor_2021}. The commercialisation of ST platforms has expanded access to these technologies, earning ST the title of “Method of the Year 2020” by Nature Methods \cite{marx_method_2021}. \\

Imaging-based fluorescence in situ hybridisation (FISH) technologies provide high-multiplex, subcellular-resolution transcriptomics data across over one million cells. These platforms, such as CosMx by NanoString and Xenium by 10x Genomics, offer high sensitivity and specificity, facilitating the exploration of cell atlases, cell–cell interactions, and the phenotypic architecture of the TME \cite{chen_spatially_2015, vandereyken_methods_2023}. \\

Despite the promise of ST, significant technical challenges remain. Two primary challenges include: (1) the integration of raw ST data into standardised and reproducible analysis workflows, which is complicated by variability in platforms and data formats; and (2) the accurate assignment of transcripts to individual cells, a task complicated by the heterogeneity and complex architecture of tissues. These challenges hinder downstream analyses such as cell type identification, spatial gene expression mapping, and inference of cell-cell interactions. Addressing these challenges is critical for fully harnessing the potential of ST. The diversity of technologies provides multiple possibilities, each with its own strengths and weaknesses, with some offering higher spatial resolution, others greater transcriptomics depth, or better compatibility with specific tissue types, features that make individual technologies better suited to answering distinct biological questions. However, a unified workflow that accommodates both platforms, from raw data processing to downstream analysis, is still needed. Establishing such a framework will streamline cross-platform data integration, unlock the full potential of spatial biology, and enable more effective multimodal analyses. \\

To address the first challenge, we developed a unified workflow that supports raw data from both CosMx and Xenium. While a Xenium reader already exists and handles multiple modalities effectively, CosMx readers lacked robustness in several areas: handling of coordinate systems, creation of segmentation polygons, and reintegration of multi-channel images. We addressed these limitations to ensure the resulting Python object matches the Xenium output format. We also aligned the workflow with the most suitable Python package, the spatialData library \cite{Marconato2025}, which integrates spatial elements (images, transcript locations, cell segmentation labels and shapes (polygons)) with transcriptomics data into an annotated dataframe suitable for single-cell analysis. \\

Addressing the second challenge requires precise spatial delineation of cells, making cell segmentation a critical step. The quality of segmentation directly affects the accuracy of all downstream analyses. Our workflow integrates native segmentation approaches: CosMx uses a Cellpose-based method \cite{Stringer2021}, while Xenium employs a Voronoi expansion strategy \cite{Janesick2023}. Users may also choose alternative or custom segmentation tools, which can offer improved performance but typically require careful parameter tuning. Such tuning is difficult to implement in tools like Xenium Ranger (10x Genomics) or AtoMx (NanoString). \\

This integrated pipeline provides a foundation for downstream modelling and analysis, offering a scalable solution for tackling key challenges in ST, especially in multimodal data integration.

\begin{figure}[h!]
  \centering
  \includegraphics[width=\linewidth]{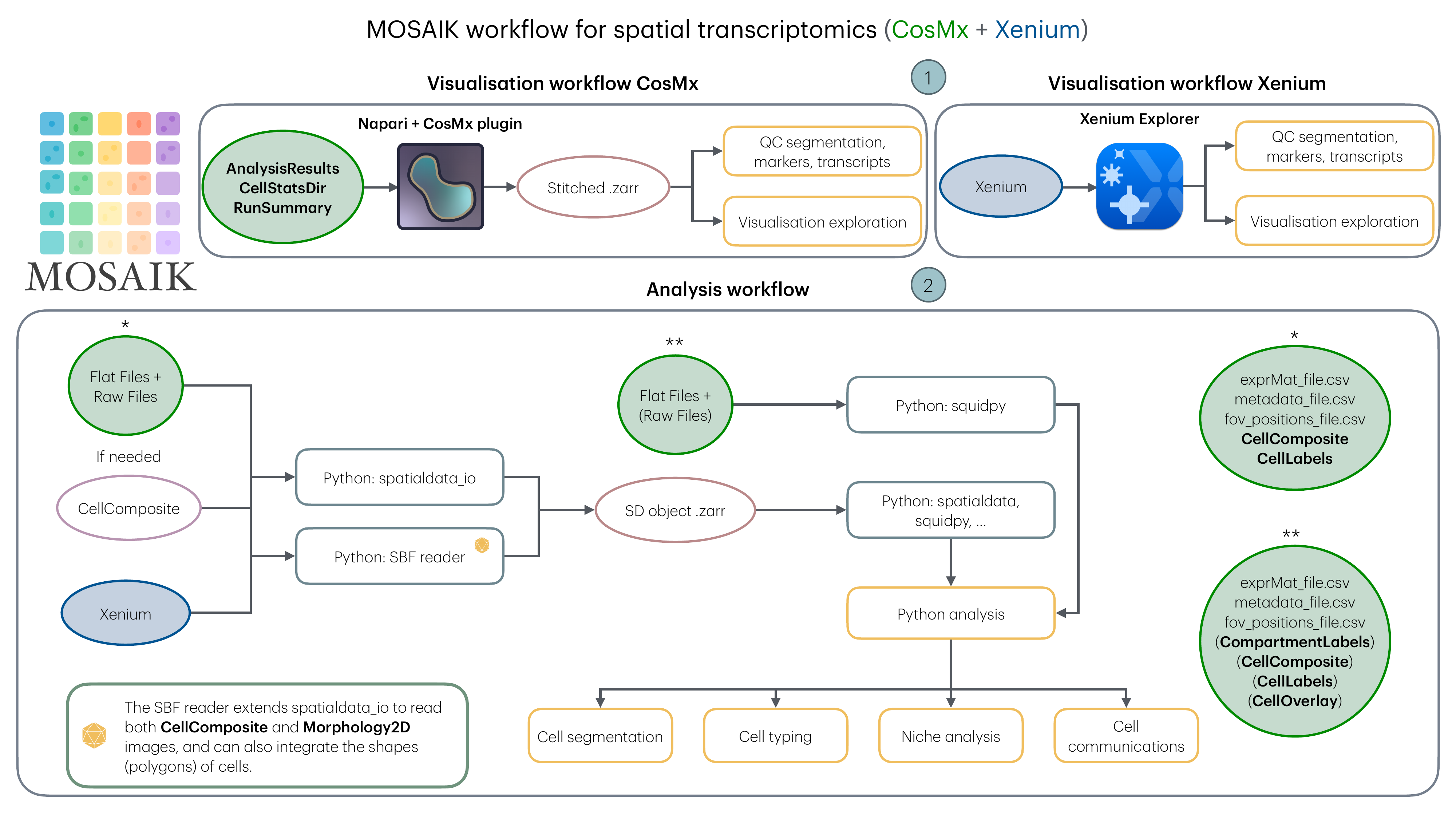}
  \caption{The MOSAIK workflow is divided into two parts: the visualisation component, which enables quality assessment of the immunofluorescence staining and verification of cell segmentation; and the data integration component, which leads to downstream analysis. 1: The MOSAIK visualisation is based on two visualisation strategies: On one hand, Napari with the CosMx plugin to visualise CosMx data; on the other hand, Xenium Explorer for Xenium data. 2: MOSAIK analysis takes the raw data and converts it into a Python object, making it easy to perform quality control and facilitate downstream analysis.}
\end{figure}

\section*{Overview of the workflow}
\hfill

The MOSAIK workflow
(\href{https://github.com/anthbapt/MOSAIK/tree/main}{https://github.com/anthbapt/MOSAIK}) supports both CosMx and Xenium ST platforms through modular pipelines designed for data integration, visualisation, and analysis (Fig. 1). For CosMx , data are first exported from the AtoMx platform, including all Flat Files and relevant Raw Files such as Morphology2D. These files are uncompressed and organised using helper scripts to generate structured directories (e.g., CellComposite,
CellLabels) essential for downstream processing. \\

Structured inputs are then read into the analysis pipeline using a custom reader, which extends the spatialdata\_io framework to incorporate various image types along with cell shape annotations (polygons). These information are stored into a Zarr file, which is open standard for storing large multidimensional array data. Then, resulting Zarr object is processed using Python-based tools such as squidpy and spatialdata for quality control and downstream analyses, including re-segmentation, cell typing, niche identification,
or cell-cell communication.  \\

Xenium data follow a similar pipeline. Data are exported directly from the instrument, processed through the same reader, and converted into a Zarr file. This unified format is then analysed using the same set of Python tools, ensuring consistency across platforms. \\

MOSAIK is the first fully integrated end-to-end workflow that supports both CosMx and Xenium raw data, standardising their output into a unified spatial data format (Fig. 2). The entire process is thoroughly documented in the \href{https://github.com/anthbapt/MOSAIK/tree/main}{MOSAIK GitHub repository}, which includes two example workflows: one using a publicly available CosMx dataset from the NanoString website, and another using a Xenium dataset from the 10x Genomics platform. \\

\begin{figure}[h!]
  \centering
  \includegraphics[width=\linewidth]{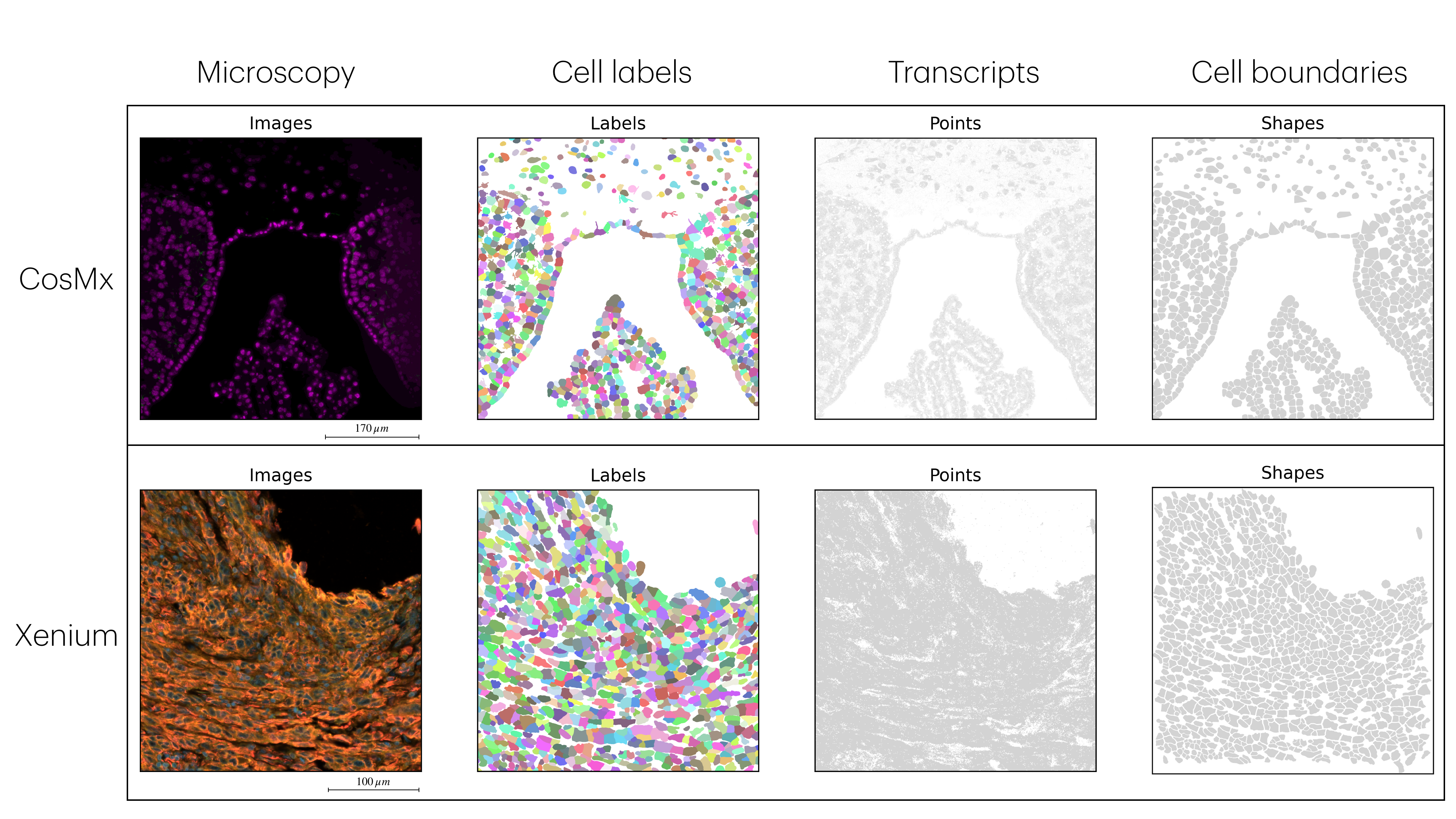}
  \caption{The Python SpatialData object obtained after using the MOSAIK workflow embeds both CosMx and Xenium data into similar objects, which can be combined or compared. MOSAIK, along with the Python library spatialdata, allows for the visualisation and connection of SpatialElements: Images (e.g., H\&E or immunofluorescence stains), Labels (segmentation maps), Points (i.e., transcripts), and Shapes (e.g., cell/nucleus boundaries or ROIs). The first two objects are raster objects (images), and the last two are vector objects (points and polygons). The CosMx fields of view are defined by a $510\,\mu\mathrm{m}$ square box, and for Xenium, each pixel represents $0.2125\,\mu\mathrm{m}$. Both CosMx and Xenium data are sourced from public repositories (see the Data availability section)}
\end{figure}

Finally, we have created a GitHub repository (\href{https://github.com/anthbapt/Spatial-Biology-Tools/tree/main}{https://github.com/anthbapt/Spatial-Biology-Tools}) that compiles a collection of Python tools designed to be used alongside or after integration with our workflow. These tools support a wide range of applications, including segmentation, cell typing, domain identification, gene imputation, detection of spatially variable genes, cell-cell communication analysis, dimensionality reduction, multimodal integration, and the use of foundation models, among others. By providing this curated collection, our goal is to guide users seamlessly from raw data to advanced analytical applications, all within a unified and community-supported framework.

\section*{Data availability}
\hfill

The datasets used to generate the figures are publicly available at the following websites: 
\begin{itemize}
    \item \href{https://nanostring.com/products/cosmx-spatial-molecular-imager/ffpe-dataset/cosmx-smi-mouse-brain-ffpe-dataset/}{https://nanostring.com/cosmx-mouse-brain-ffpe}
    \item \href{https://www.10xgenomics.com/datasets/xenium-prime-ffpe-human-skin}{https://www.10xgenomics.com/xenium-prime-ffpe-human-skin}
\end{itemize}
The processed datasets associated with the code are provided as examples in the following Zenodo repository \href{https://doi.org/10.5281/zenodo.15365593}{https://doi.org/10.5281/zenodo.15365593}.

\section*{Code availaility}
\hfill

The MOSAIK workflow is publicly available on GitHub at \href{https://github.com/anthbapt/MOSAIK/tree/main}{https://github.com/anthbapt/MOSAIK}.

\section*{Related software}
\hfill

This work integrates nicely with the existing ST community, particularly the tools that are part of the
\href{https://scverse.org}{scverse ecosystem}.

\section*{Planned enhancements}
\hfill

Recognising that ST is a rapidly evolving field,
MOSAIK is designed to remain aligned with the latest standards, both in terms of experimental setup and raw data processing, as well as on the computational side by integrating emerging methods and developmental
tools. As part of the King's College London Spatial Biology Facility (SBF), MOSAIK must stay up to date to help the SBF fulfil its mission. \\

Furthermore, newly developed tools within the group will be directly integrated into MOSAIK. This will provide the broader community with the ability to use both their own methods and those developed by our team,
methods that have been tested across a wide range of tissue types and technologies, thanks to the strong network surrounding the facility. \\

The tools that will be natively integrated into MOSAIK include segmentation methods based on SAM, as well as a multimodal integration approach that combines transcriptomics and spatial information to generate a more robust latent representation. The current modalities under consideration include H\&E, Akoya PhenoCycler, IMC, and metallomics data.

\section*{Data Acknowledgements}
\hfill

Anthony Baptista, and Anita Grigoriadis acknowledge support from the CRUK City of London Centre Award [CTRQQR-2021/100004]. Anthony Baptista, Rosamond Nuamah, Ciro Chiappini, and Anita Grigoriadis acknowledge support from MRC [MR/X012476/1]. 

\bibliographystyle{unsrt}
\bibliography{paper}

\end{document}